\documentclass[12pt]{article}
\usepackage[nottoc,notlot,notlof]{tocbibind}
\usepackage{amsfonts}
\usepackage{graphicx}
\usepackage{amsmath,amssymb}
\usepackage{epsfig}
\usepackage{dsfont}
\usepackage{color}
\usepackage{xcolor}
\usepackage{mathrsfs}
\usepackage{subfig}
\usepackage{enumerate}
\usepackage{bbm}
\usepackage{ragged2e}
\usepackage{tabularx}
\usepackage{float}
\usepackage{adjustbox}
\usepackage{booktabs}
\usepackage{chngpage}
\usepackage[flushleft]{threeparttable}

\newcommand{\blind}{1}

\def\nbgroup{N}

\def\temps{T_{i}}
\def\censure{C_{i}}
\def\yobs{X_{i}}
\def\bX{\mathbf{X}}

\def\dobs{\Delta_{i}}

\def\hazard{h_{i}(t|b_i)}

\def\baseline{h_0(t)}

\def\desfixe{Z_{i}}

\def\fixe{\beta}

\def\frailty{b_i}
\def\frailtyv{\mathbf{ b}}

\def\exp{\text{ exp}}

\def\Sborel{\mathcal{S}}
\def\Sset{\text{S}}

\DeclareMathAlphabet\mathbfcal{OMS}{cmsy}{b}{n}

\def\1{1\!{\mathrm l}}

\def\bdelta{\mathbf{\Delta}}

\newcommand{\norm}[1]{\left\lVert#1\right\rVert}

\usepackage[numbers]{natbib}
\setcitestyle{authoryear,open={(},close={)}}

\newtheorem{theorem}{Theorem}

\begin{document}

\def\spacingset#1{\renewcommand{\baselinestretch}%
{#1}\small\normalsize} \spacingset{1}


\if1\blind
{
  \title{Modeling dependent survival data through random effects with spatial correlation at the subject level}
  \author{Ajmal Oodally \\
    MaIAGE, Universit\'e Paris-Saclay, INRAE, France \\
    and \\
    Estelle Kuhn \\
    MaIAGE, Universit\'e Paris-Saclay, INRAE, France \\
    and \\
    Klara Goethals \\
    Faculty of Veterinary Medicine, Ghent University, Belgium \\
    and \\
    Luc Duchateau \\
    Faculty of Veterinary Medicine, Ghent University, Belgium
    }
  \maketitle
} \fi

\if0\blind
{
  \bigskip
  \bigskip
  \bigskip
  \begin{center}
    {\LARGE \bf Modeling dependent survival data through random effects with spatial correlation at the subject level  
 \par}
\end{center}
  \medskip
} \fi

\bigskip
\begin{abstract}
Dynamical phenomena such as infectious diseases are often investigated by following up subjects longitudinally, thus generating time to event data. The spatial aspect of such data is also of primordial importance, as many infectious diseases are transmitted from one subject to another. In this paper, a spatially correlated frailty model is introduced that accommodates for the correlation between subjects based on the distance between them. Estimates are obtained through a stochastic approximation version of the Expectation Maximization algorithm combined with a Monte-Carlo Markov Chain, for which convergence is proven. The novelty of this model is that spatial correlation is introduced for survival data at the subject level, each subject having  its own frailty. This univariate spatially correlated frailty model is used to analyze spatially dependent malaria data, and its results are compared with other standard models.        
\end{abstract}

\noindent%
{\it Keywords:}  survival analysis, frailty model, correlated frailties, time to malaria infection, expectation-maximization algorithm, stochastic approximation, Monte-Carlo Markov chain method. 
\vfill

\newpage
\spacingset{1.5} 
\section{Introduction}
\label{sec:intro}


Malaria remains a disease with high morbidity and mortality in Ethiopia according to the most recent information from the \cite{worldmalaria2018}, with about 74 million people at risk for the disease. Additionally, Ethiopia launched an ambitious green energy program making use of the large altitude differences in the country to generate electricity from hydroelectric dams. Such dams, however, provide excellent breeding grounds for the vector of the malaria parasite, the Anopheles mosquito, and thus might lead to increasing malaria incidence.
In order to investigate the dam effect, data have been collected in villages around the Gilgel-Gibe hydroelectric dam in the Oromia region at different distances from the dam. \cite{getachew2013} reported the results of this study using a shared frailty model. In this paper, we revisit these data and develop new survival models to better mimic the correlation structure in the data. The use of spatial statistic tools is relatively new in survival analysis although it may substantially improve dependent survival data modelling. \cite{banerjee2003frailty} proposed a parametric frailty model to estimate parameters using a Bayesian approach to analyse an infant mortality dataset in Minnesota. Spatial dependence between the clusters was modelled using two different approaches; a geostatistical approach where the exact locations are needed and a lattice approach where the relative distance between the groups is required. \cite{louise02} developed a semi-parametric spatial frailty model with Monte Carlo simulations and Laplace approximation of a rank based marginal likelihood. Along the same lines, \cite{lin2012} estimated parameters of a log-normal spatial frailty model using a two-iteration approach based on an approximate likelihood function, alternating between the estimation of the regression parameter and the variance components. 

However, all of the above models introduced spatial correlation between groups of subjects. In our approach, we go beyond that and model spatial correlation at the subject level. In Section 2, the model is detailed followed by the estimation procedure in Section 3. Simulation results are presented in Section 4 and the analysis of the malaria data set is discussed in Section 5. Conclusion and discussion  follow in Section 6.

\section{Description of the univariate spatially correlated frailty model}
\label{sec:spatial_frailty_model}
We propose a univariate frailty model with spatial correlations at the subject level. Let us consider  $N$ subjects. For $1 \leq i \leq \nbgroup$, the time to event and the time of censoring for subject $i$ are modelled by random variables denoted by $\temps$ and $\censure$ respectively. Then, for $1 \leq i \leq N$, the right censored time and the censoring indicator are denoted by $\yobs$ and $\dobs$ respectively and defined by $\yobs = \text{min} (\temps, \censure)$ and $\dobs = \mathbbm{1}_{\temps \leq \censure}$. We denote $\bX = (\yobs)_{1 \leq i \leq N}$ and $\bdelta = (\dobs)_{1 \leq i \leq N}$ . The spatially correlated univariate frailty model is defined as follows. For $1 \leq i \leq N$ the conditional instantaneous hazard of occurrence of the event for subject
$i$ at time $t$ is defined by:

\begin{equation}
\label{model:spatial_model}
\hazard = \baseline \ \exp (\desfixe^t \fixe + \frailty)
\end{equation}

where $\baseline$ is the baseline hazard function at time $t$, $\frailty$ the frailty term of subject $i$, $\fixe$ the vector of the unknown regression parameters, $\desfixe$ the vector of covariates associated with subject $i$. Let us introduce the frailty vector $\frailtyv = (\frailty)_{1 \leq i \leq \nbgroup}$ which is assumed to follow a centered multivariate normal distribution with covariance matrix $\sigma^2 \Sigma (\rho)$ where $\sigma^2 $ is a scaling factor and $\Sigma(\rho)$ is the correlation matrix parameterized by $\rho > 0$, 
\begin{equation}
\label{model:frailty}
\frailtyv \sim \mathcal{N} (0,\sigma^2 \Sigma (\rho)).
\end{equation}
We consider two different correlation structures following \cite{louise02}, namely $\Sigma_{\text{exp}}(\rho) = \text{exp} (-\rho D)$ and $\Sigma_{\text{pol}}(\rho) = \frac{1}{1 + D^{\rho}}$ with $D = (d_{ii^{'}}) \in \mathcal{M}_N(\mathbb{R}^{+})$ where $d_{ii^{'}}$ corresponds to the distance between subject $i$ and subject $i^{'}$. By definition, $d_{ii}=0$.

For the baseline hazard function, usual regular parametric forms can be considered. Some examples include the Weibull, Gompertz and piecewise constant baselines. We denote by $\alpha$ the parameters of the baseline hazard function $h_0$. 

Finally, the whole model parameters are $\theta = (\alpha, \beta, \sigma^2, \rho)$.

\section{Estimation in the spatially correlated frailty model}
\label{sec:estimation_method}

\subsection{Definition of the maximum marginal likelihood estimate}
We estimate the parameters of the model by maximising the marginal likelihood. We introduce  the following assumptions on the univariate spatially correlated frailty model: 
\justify
\textbf{(F1)} The  censoring times $(\censure)$ are independent of the event times $(\temps)$ and of the frailty vector $\frailtyv$. \\
\textbf{(F2)} Conditional on the frailty vector $\frailtyv$, the event times $(\temps)$ are independent.

By assumptions \textbf{(F1)} and \textbf{(F2)}, the complete likelihood can be expressed as:

\begin{align}
\label{eq:complete_likelihood_spatial}
L_{\text{comp}} (\theta;\bX,\bdelta,\frailtyv) = \prod_{i=1}^{N} \Bigg( \frac{(h_0 (\yobs) \exp(\desfixe^t \beta + \frailty) )^{\dobs}}{\exp(H_0(\yobs) \exp(\desfixe^t \beta + \frailty))} \Bigg) f(\frailtyv)
\end{align}

\sloppy
where $H_0(\yobs) = \int_0^{\yobs} h_0 (t) dt$ is the cumulative hazard function and $f$ is the density of a centered multivariate gaussian distribution with covariance matrix $\sigma^2 \Sigma (\rho)$.  

The marginal likelihood is obtained by integrating the complete likelihood with respect to the frailty vector $\frailtyv$: 

\begin{equation*}
L_{\text{marg}} (\theta;\bX,\bdelta) = \int L_{\text{comp}}(\theta;\bX,\bdelta,\frailtyv) d\frailtyv
\end{equation*}

The estimator of the maximum of the marginal likelihood $\hat{\theta}$ is defined by:
$$\hat{\theta} = \text{argmax} \ L_{\text{marg}} (\theta;\bX,\bdelta).$$ 
 This estimator cannot be evaluated directly in many models, in particular when the marginal likelihood does not admit an analytical form, which is the case in the frailty model we consider. In practice, we calculate the value of the estimator using an iterative algorithm. 

\subsection{Computation of the estimate using a stochastic algorithm}
\label{section:saem_mcmc_adaptative_gibbs_block}

We apply the SAEM-MCMC algorithm with truncation on random boundaries (\cite{KuhnLavielle04}, \cite{stephkuhn}) to compute the maximum of the marginal likelihood. Let us describe in detail this algorithm, denoted later by Algorithm $\mathcal{A}$. First note that the complete likelihood function defined in equation (\ref{eq:complete_likelihood_spatial}) belongs to the  exponential family since it can  be written as follows:
\begin{equation}
\label{eq:Lcomp_spatial_exponential}
 L_{\text{comp}}(\theta;\bX,\bdelta,\frailtyv) =  \text{exp} (-\Psi(\theta) + \langle \Sborel (\frailtyv), \Phi(\theta) \rangle )
\end{equation}
\sloppy where $\Sborel$, $\Psi$ and $\Phi$ are  Borel functions. Sufficient statistics can be expressed as $\Sborel(\frailtyv) = \Big(  b_{i} b_{i^{'}}, i,i' = 1,\dots, N, \exp(b_{i}), i = 1,\dots, N \Big)^t$ and take values in a subset $\Sset$ of $\mathbb{R}^{N(N+3)/2}$.

Let $(\mathcal{K}_{q})_{q} \geq 0$ be a sequence of increasing compact subsets of $\Sset$ such that $\bigcup_{q \geq 0} \mathcal{K}_{q} = \Sset$ and $\mathcal{K}_q \subset \text{int}(\mathcal{K}_{q+1})$, for all $q \geq 0$. Initialize $\theta_0$ in $\Theta$, $\frailtyv_0$ and $s_0$ in two fixed compact sets $\textbf{K}$ and $\mathcal{K}_0$ respectively. Each iteration of Algorithm $\mathcal{A}$ is composed of four steps detailed below. 

Repeat until convergence for $k \geq 1$:

\begin{enumerate}
\item \textbf{Simulation step:} Draw a realization $\bar{\frailtyv}$ of the unobserved frailty vector from a transition probability $\Pi_{\theta}$ of a convergent Markov chain having as stationary distribution the conditional distribution $\pi_{\theta} (. | \bX,\bdelta)$ defined by $\pi_{\theta} (\frailtyv | \bX,\bdelta) = L_{\text{comp}} (\theta;\bX,\bdelta,\frailtyv) \mathbin{/} L_{\text{marg}} (\theta;\bX,\bdelta)$ with the current parameters:
 $$\bar{\frailtyv} \sim \Pi_{\theta_{k-1}} (\frailtyv_{k-1},.)$$
\item \textbf{Stochastic approximation step:} Compute $\bar{s} = s_{k-1} + \mu_{k} (\Sborel(\bar{\frailtyv}) - s_{k-1})$ 
where the sequence $(\mu_k)_k$ satisfies $0 \leq \mu_k \leq 1,  \sum \mu_k = +\infty,  \sum \mu_k^2 < +\infty$.
\item \textbf{Truncation step:} If $\bar{s}$ is outside the current compact set $\mathcal{K}_{\kappa_{k-1}}$, where $\kappa$ is the index of the current active truncation set, or too far from the previous value $s_{k-1}$ then restart the stochastic approximation in the initial compact set, extend the truncation boundary to $\mathcal{K}_{\kappa_{k}}$ and start again with a bounded value of the missing variable. Otherwise if $\norm{s_k - s_{k-1}} \leq \epsilon_{k}$ where $\epsilon = (\epsilon_k)_{k \geq 0}$ is a monotone non-increasing sequence of positive numbers, set $(\frailtyv_k,s_k) = (\bar{\frailtyv},\bar{s})$ and keep the truncation boundary to $\mathcal{K}_{\kappa_{k-1}}$. 
\item \textbf{Maximization step:} 
    \begin{equation*}
        \theta_k = \underset{s}{\text{argmax }} \hat{\theta}(s)
    \end{equation*}
    where the function $\hat{\theta}: \Sset \rightarrow \Theta$ is defined such that:
$$\forall \ s \in \Sset, \forall \ \theta \in \Theta, \ L(\hat{\theta}(s),s) \geq L(\theta,s)$$
with $L(\theta,s) = \text{exp} (- \Psi (\theta) + \langle s, \Phi (\theta) \rangle )$. 
\end{enumerate}

Note that the truncation step guarantees two conditions on the sequence $(\frailtyv_k,s_k)$ generated by this algorithm at each iteration $k$. Namely it ensures that the stochastic approximation does not wander outside the current compact set and that the current value is not too far from the previous value. Indeed the truncation step is introduced following \cite{moulines05} to allow for frailties that do not live in a compact set and thereby establish the convergence proof with weak assumptions on the model.

\subsection{Convergence property of Algorithm $\mathcal{A}$}
In this section, we prove the almost sure convergence of the sequence $(\theta_k)_k$ generated by Algorithm $\mathcal{A}$ to a critical point of the marginal likelihood. Let us first detail the assumptions made on the model, the dynamic of the algorithm and the Markov kernel in the simulation step. 
\justify
First, we make classical model assumptions \textbf{(F3)}-\textbf{(F5)}, corresponding to  assumptions \textbf{(M3)}-\textbf{(M5)} of \cite{dely99} and detailed in Appendix \ref{appendix:proof_theorem}, to prove the convergence of EM like algorithms following those of \cite{dely99}. Then, following \cite{moulines05}, we consider a global Lyapunov function denoted by $w$ defined as:

\begin{equation}
\label{eq:w_s_spatial}
    w(s) = - \log \int L_{\text{comp}} (\hat{\theta} (s); \bX,\bdelta,\frailtyv) d\frailtyv
\end{equation} 

and a mean field $h$ defined as: 

\begin{equation}
\label{eq:h_s_spatial}
    h(s) = \int (\Sborel(\frailtyv) - s) \pi_{\hat{\theta}(s)} (\frailtyv | \bX,\bdelta) d\frailtyv
\end{equation}

We state a first assumption on these functions.
\justify
\textbf{(F6)} The functions $w$ and $h$ are such that
\begin{enumerate}[\hspace{18pt}(i)]
\item there exists an $M_0 > 0$ such that 
    \begin{equation*}
      \Sset_0= \lbrace s \in \Sset, \langle \nabla w(s),h(s)\rangle = 0 \rbrace \subset \lbrace s \in \Sset, w(s) < M_0 \rbrace
    \end{equation*}
    where $w$ is defined in (\ref{eq:w_s_spatial}) and $h$ is defined in (\ref{eq:h_s_spatial}).
\item there exists $M_1 \in ]M_0,\infty]$ such that $\lbrace s \in \Sset, w(s) < M_1 \rbrace$ is a compact set.
\item the closure of $w(\mathcal{L})$ has an empty interior.
\end{enumerate}

We now state a condition on the step-size sequences in the stochastic approximation and truncation steps of the algorithm.
\justify
\textbf{(F7)} The sequences $\mu = (\mu_k)_{k \geq 0}$ and $\epsilon = (\epsilon_k)_{k \geq 0}$ are non-increasing, positive and satisfy $\sum_{k=0}^{\infty} \mu_k = \infty$, $\underset{k \rightarrow \infty}{\text{lim}} \epsilon_k = 0$ and $\sum_{k=1}^{\infty} \lbrace \mu_k^2 + \mu_k \epsilon_k^a + (\mu_k \epsilon_k^{-1})^p \rbrace < \infty$, where $a \in ]0,1]$ and $p \geq 2$. 
\justify
Assumptions  \textbf{(F6)} and \textbf{(F7)} correspond to assumptions \textbf{(A1)} and \textbf{(A4)} of \cite{moulines05} respectively. Finally we consider the usual drift assumption \textbf{(DRI)} classical in Markov chain literature detailed in \cite{moulines05}. 

\begin{theorem}
\label{thm:adaptative_SAEM_MCMC}
Assume that \textup{\textbf{(F1)-(F7)}} and \textup{\textbf{(DRI)}} are fulfilled. Then we have with probability 1 
\begin{equation*}
    \underset{k \rightarrow \infty}{\text{lim}} d(\theta_k, \mathcal{L}) = 0
\end{equation*}
where $(\theta_k)_k$ is generated by Algorithm $\mathcal{A}$, $d(x,A)$ denotes the distance from $x$ to any closed subset $A$  and $\mathcal{L} = \lbrace \theta \in \Theta, \partial_{\theta} \text{ log } L_{\text{marg}} (\theta; \bX, \bdelta) = 0 \rbrace$.
\end{theorem}

We refer to Appendix \ref{appendix:proof_theorem} for the proof of Theorem \ref{thm:adaptative_SAEM_MCMC}. 

\subsection{Practical details on the implementation of Algorithm $\mathcal{A}$}

\begin{enumerate}
\item In order to reduce the computational time to achieve convergence, it is crucial to start the algorithm with good initial estimates. The usual approach is to start with estimates obtained from simpler existing methods. For instance, we use as initial values for the regression parameter $\beta$ and baseline components the estimated values obtained when fitting the data by a piecewise constant proportional hazards model. 

\item At iteration $k$ of the algorithm, the transition kernel used for simulating the unobserved frailty is often chosen as a transition kernel of a random walk Metropolis-Hastings algorithm where the proposal distribution $q$ equal to a normal distribution centred at the current value $\frailtyv_{k-1}$. However, if the frailty vector $\frailtyv$ is of high dimension, this can be inefficient in practice. To cope with this high dimensionality, one can implement a hybrid Gibbs algorithm where $\frailtyv_{k-1}$ is updated coordinate-wise,  each candidate being accept with an acceptance probability that has to be computed $N$ times for a frailty vector of size $N$. Nevertheless this step can be computationally intensive with an increasing size $N$ of the frailty vector. Instead, one can update blocks of size $K$ at a time. In doing so, the computational cost is reduced by a factor of $K$. A subtle compromise has to be found between a too big value of $K$ which translates to a change in too many directions at a time for $\frailtyv$ leading to a possible extremely low acceptance rate and a too small $K$ leading to a high computational cost. Moreover we recommend to use an adaptive version to achieve reasonable acceptance rates (\cite{haario2001adaptive}).

\item The decreasing positive step size $(\mu_k)_k$ is taken as follows for all $0 \leq k \leq  K_0, \ \mu_k = 1$ and for all $k > K_0,  \  \mu_k = \frac{1}{(k-K_0)} $  
   where  $K_0$ is a number to be specified. The step size $(\mu_k)_k$ verifies assumption \textbf{(F7)}. The algorithm is said to have no memory during the first  $K_0$ iterations. After this burn-in time which allows for the algorithm to widely visit the parameter space, the sequence $(\mu_k)_k$ decreases and converges to zero as $k \rightarrow \infty$. Besides, there is no need  to implement in practice the truncation step which is only required for the theoretical proof of convergence. Thus we state no recommendation for   the sequence $(\epsilon_k)_k$.
   \item Following \cite{ripatti2002maximum}, we consider a stopping criterion based on the relative difference between two consecutive values of the parameters. Let us fix a positive  threshold $\epsilon > 0$. If for some $k>1$:
    \begin{equation*}
        \frac{\norm{\theta_k - \theta_{k-1}}}{\norm{\theta_{k-1}}} < \epsilon
    \end{equation*}
    holds true for  three  consecutive iterations, the algorithm is stopped. 
\end{enumerate}

All programs are available on request from the authors.

\section{Simulation study}

\subsection{Study of the consistency of the estimate $\hat{\theta}$}

The simulation setting is chosen to mimic the malaria data (\cite{getachew2013}). We generate event times for $N = 300$ subjects using model $(\mathbfcal{M}_1)$ which is defined as follows: 
\begin{equation}
\tag{$\mathbfcal{M}_1$}
\label{model:M1}
\forall 1 \leq i \leq N \ \hazard = \sum_{m=1}^3 h_m  \mathbbm{1}_{ [ \tau_{m-1}, \tau_m [ } (t) \exp (\desfixe^t \fixe + \frailty) \ \ \text{with} \ \frailtyv \sim \mathcal{N}(0,\sigma^2 \Sigma_{\text{exp}}(\rho))
\end{equation} 
where the baseline hazard is piecewise constant with change points $(\tau_0,\tau_1,\tau_2,\tau_3) = (0,0.2,0.8,+\infty)$ and constant hazards $(h_1,h_2,h_3) = (2,0.5,1)$. The parameter vector $\beta$ is set equal to (2,3) and the covariates $(Z_i)$ are simulated following a Bernoulli distribution of parameter $0.5$. The parameters of the correlation structure are fixed at $\sigma^2 = 1.5$ and $\rho = 1$. The matrix $D$ is chosen by taking subsets of size $300$ of the real malaria distance matrix. We simulate the event times under three different censoring settings namely no censoring, moderate censoring ($40 \%$) and heavy censoring ($60 \%$). The censoring times are simulated following an exponential distribution with the rate parameter adjusted so as to obtain the desired censoring level.

\begin{table}[H]
\begin{center}
\caption[Numerical consistency of univariate spatially correlated frailty model estimates]{Mean of the parameter
estimates and empirical standard error in parentheses estimated in model \eqref{model:M1} from $100$ repetitions with data generated under model \eqref{model:M1}. The number of subjects $N$ is fixed at $300$. \label{tab:consistency_estimaes_spatial_model}}
\begin{tabular}{|l|l|l|l|l|l|l|}
\hline
Parameters & True Values & No censoring & 40 $\%$ censoring & 60 $\%$ censoring \\
\hline
\hline
$h_1$ & 2 & 1.942 (0.961)  & 2.146 (1.106) & 2.043 (1.124) \\
$h_2$ & 0.5 & 0.473 (0.259)  & 0.521 (0.296) & 0.488 (0.290) \\
$h_3$ & 1 & 0.957 (0.447)  & 1.089 (0.611) & 1.209 (0.884) \\
\hline
$\beta_1$ & 2 & 2.001 (0.170) & 2.013 (0.206) & 2.002 (0.292) \\
$\beta_2$ & 3 & 2.969 (0.210) & 3.010 (0.254) & 3.061 (0.340) \\
\hline
$\sigma^2$ & 1.5 & 1.554 (0.444) & 1.642 (0.463) & 1.654 (0.552)\\
$\rho$ & 1 & 0.977 (0.277) & 1.051 (0.318) & 1.072 (0.322) \\ 
\hline
\end{tabular}
\end{center}
\end{table}

The estimates are computed using Algorithm $\mathcal{A}$ and presented in Table \ref{tab:consistency_estimaes_spatial_model}. All estimates are close to the true values, whatever the censoring setting. We emphasize that the lower the censoring rate, the closer the estimates to the true values are. We also note that the standard errors are larger with increasing proportions of censored data for all estimates. For instance, the standard error for $\hat{h}_3$ nearly doubles when we compare the non-censoring setting to heavy censoring with respective standard errors of $0.447$ and $0.884$.

\subsection{Robustness to misspecification of the correlation structure}
\label{simulation:robustness_misspecification_correlation}

To evaluate effects of misspecification with respect to correlation structure, we introduce model $(\mathbfcal{M}_2)$ defined by:
\begin{equation}
\tag{$\mathbfcal{M}_2$}
\label{model:M2}
\forall 1 \leq i \leq N \ \hazard = \sum_{m=1}^3 h_m  \mathbbm{1}_{ [ \tau_{m-1}, \tau_m [ } (t) \exp (\desfixe^t \fixe + \frailty) \ \ \text{with} \ \frailtyv \sim \mathcal{N}(0,\sigma^2 \Sigma_{\text{pol}}(\rho))
\end{equation} 

We fit the model \eqref{model:M2} for data simulated under the model \eqref{model:M1}. The results are shown in Table \ref{tab:robustness_misspecification_correlation_structure}.

\begin{table}[H]
\begin{center}
\caption[Parameter estimates : robustness with respect to misspecification of the correlation structure]{Mean of the parameter estimates and empirical standard error in parentheses estimated in model \eqref{model:M2} from 100 repetitions with data generated under model \eqref{model:M1}. The number of subjects $N$ is fixed at 300. \label{tab:robustness_misspecification_correlation_structure}}
\begin{tabular}{|l|l|l|l|l|}
\hline
Parameters & True Values & No censoring & 40 $\%$ censoring& 60 $\%$ censoring \\
\hline
\hline
$h_1$ & 2 & 2.276 (1.600) & 2.298 (1.642) & 2.556 (2.223) \\
$h_2$ & 0.5 & 0.582 (0.440) & 0.591 (0.486) & 0.680 (0.621) \\
$h_3$ & 1 & 1.256 (0.839) & 1.286 (0.893) & 1.595 (1.982) \\
\hline
$\beta_1$ & 2 & 2.048 (0.186) & 2.025 (0.216) & 2.038 (0.275) \\
$\beta_2$ & 3 & 3.098 (0.223) & 3.045 (0.276) & 3.086 (0.333) \\
\hline
$\sigma^2$ & 1.5 & 1.932 (0.495) & 1.805 (0.566) & 1.946 (0.590) \\
$\rho$ & 1 & 0.817 (0.164) & 0.748 (0.168) & 0.648 (0.167) \\
\hline
\end{tabular}
\end{center}
\end{table}

The estimates of the regression parameters $\beta$ are quite robust to the misspecification of the covariance structure. However, the estimates of the baseline parameters $(h_m)_{1 \leq m \leq 3}$ are far from the true values and even more so in the case of moderate and heavy censoring. When the correlation structure is misspecified, the scaling factor $\sigma^2$ seems to be compensating for the wrong assumption of correlation structure which leads to an overestimation of this parameter.

\subsection{Comparison with other models without spatial correlation structure}
We compare the estimator of the univariate spatially correlated frailty model with two existing models that do not take into account spatial correlation based on the same simulation setting. We define the proportional hazards model \eqref{model:M3} and the univariate frailty model \eqref{model:M4} as follows:

\begin{equation}
\tag{$\mathbfcal{M}_3$}
\label{model:M3}
\forall 1 \leq i \leq N \ \ \hazard = \sum_{m=1}^3 h_m  \mathbbm{1}_{ [ \tau_{m-1}, \tau_m [ } (t) \exp (\desfixe^t \fixe)
\end{equation} 

\begin{equation}
\tag{$\mathbfcal{M}_4$}
\label{model:M4}
\forall 1 \leq i \leq N \ \ \hazard = \sum_{m=1}^3 h_m  \mathbbm{1}_{ [ \tau_{m-1}, \tau_m [ } (t) \exp (\desfixe^t \fixe + \frailty) \ \text{with} \ \ \frailtyv \sim \mathcal{N}(0,\sigma^2)
\end{equation}

Neither the proportional hazards model \eqref{model:M3} nor the univariate frailty model \eqref{model:M4} accommodate for spatial correlation in the data. The estimates are far from the true values in both cases as shown in Table \ref{tab:comparing_coxme_phreg}. The variance of the frailties in model \eqref{model:M4} is large at a value of 0.988, and thus overdispersion seems to be present in the data. This adjustment makes that the estimates of model \eqref{model:M4} are closer to the true values than estimates of model \eqref{model:M3}. Similar results are obtained for different censoring settings (results not presented).

\begin{table}[H]
\begin{center}
\caption[Comparison of different estimators for simulated spatially correlated data]{Mean of the parameter
estimates and empirical standard error in parentheses estimated in model \eqref{model:M3} and model \eqref{model:M4} from $100$ repetitions with data generated under model \eqref{model:M1}. The number of subjects $N$ is fixed at $300$. \label{tab:comparing_coxme_phreg}}
\begin{tabular}{|l|l|l|l|l|}
\hline
Parameters & True Values & Proportional hazards model & Univariate frailty model \\
\hline
\hline
$h_1$ & 2 & 2.583 (0.721) & 2.172 (0.688) \\
$h_2$ & 0.5 & 0.351 (0.128) & 0.455 (0.194) \\
$h_3$ & 1 & 0.298 (0.115) & 0.757 (0.342) \\
\hline
$\beta_1$ & 2 & 1.555 (0.210) & 1.874 (0.250) \\
$\beta_2$ & 3 & 2.299 (0.269) & 2.835 (0.294) \\
\hline
$\sigma^2$ & 1.5 & $\times$ & 0.988 (0.270) \\
\hline
\end{tabular}
\end{center}
\end{table}

\section{Analysing the Gilgel Gibe time to malaria data set}

\subsection{Data description}
\begin{figure}[!httpb]
    \centering
    \includegraphics[scale=0.4]{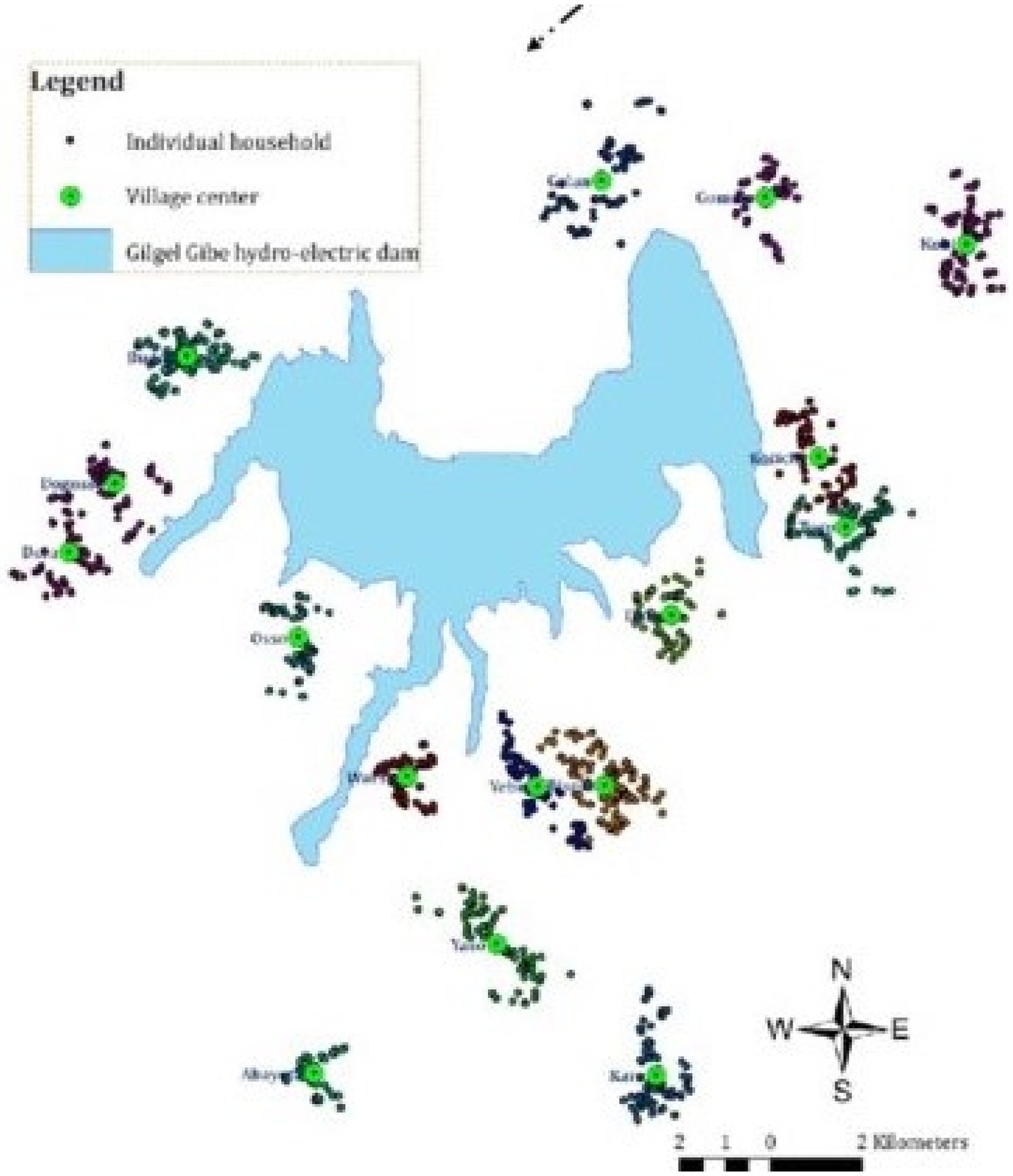}
    \caption{Gilgel-Gibe hydroelectric dam, study villages and households (\cite{getachew2013})}
    \label{fig:gibel_dam}
\end{figure}

In the Gilgel Gibe study, 2037 children living in 16 different villages were followed over time (\cite{yewhalaw2013}). The location of the children is presented in Figure \ref{fig:gibel_dam}, which demonstrates that the village boundaries are mostly of an administrative nature, and that villages are almost overlapping in some cases. The geographical coordinates (longitude and latitude) of the children around the dam are used to compute the inter-distances between pairs of children. We consider four covariates, most importantly, the distance to the dam and next to that the sex of the child, the age and the structure of the roof of the child's household. The children are grouped into three age categories namely, 0-3 (used as reference group), 3-7 and older than 7. Some children have exactly the same geographical coordinates (same household) or live so close to each other that the locations recorded are exactly the same. We assign a common frailty term to those children so that the distance matrix remains invertible and positive definite. We note however that this grouping structure is different from the usual grouping structure in spatial survival models in the sense that the members of the group have exactly the same geographical coordinates. In the common grouping structure, the group normally refers to a region, state or country (\cite{louise02}, \cite{banerjee2016spatial}). The number of children in a group varies from 1 to 11 with most groups consisting of 1, 2 or 3 children (760 groups of 1, 364 groups of 2 and 89 groups of 3).

\subsection{The univariate spatially correlated frailty model}
\label{sec:malaria_data_analysis}

We fit the univariate spatially correlated frailty model defined in equations \eqref{model:spatial_model} and \eqref{model:frailty} to the data. The parametric baseline hazard is chosen to be piecewise constant, using the rainfall data to determine adequate cut-points as described in \cite{belay2017}. The parameters of the model are estimated using the SAEM-MCMC algorithm detailed in Section \ref{section:saem_mcmc_adaptative_gibbs_block} with a Gibbs-block of size $K = 10$. Smaller block sizes give similar results but need longer computational times. Besides, the algorithm fails to converge for larger block sizes. Initial parameter estimates for $\beta$ and the baseline components are obtained using the R package \textit{eha} (\cite{brostrom2012eha}) which allows for estimation in a piecewise baseline proportional hazards model. The marginal log-likelihood values for a model with spatial correlation structure given by $\Sigma_{\text{pol}}(\rho)$, equal to -6013.608, is substantially larger than the model with spatial correlation structure given by $\Sigma_{\text{exp}}(\rho)$, equal to -6038.579. Therefore, we proceed with the former model to investigate the spatial correlation structure and the covariate effects. 

We also investigate whether the correlation term $\rho$ is significantly different from $\infty$, i.e. whether the univariate spatially correlated frailty model is a better fit than a univariate frailty model with no spatial correlation. We perform a likelihood-ratio test with null hypothesis ``$\rho = \infty$". We note that the parameter tested here is on the boundary of the parameter space. Therefore, following the works of \cite{self1987asymptotic}, the likelihood-ratio test statistic converges asymptotically to a 50:50 mixture of $\delta_0$ and $\chi^2(1)$ distributions if the null hypothesis is correct. With $Y$ a random variable from this mixture distribution, the p-value is given by $\mathbb{P} (Y > 7.64) = 0.003$ and we can thus reject the null hypothesis at the $1 \%$ significance level.

\begin{figure}[H]
    \centering
    \includegraphics[scale=0.7]{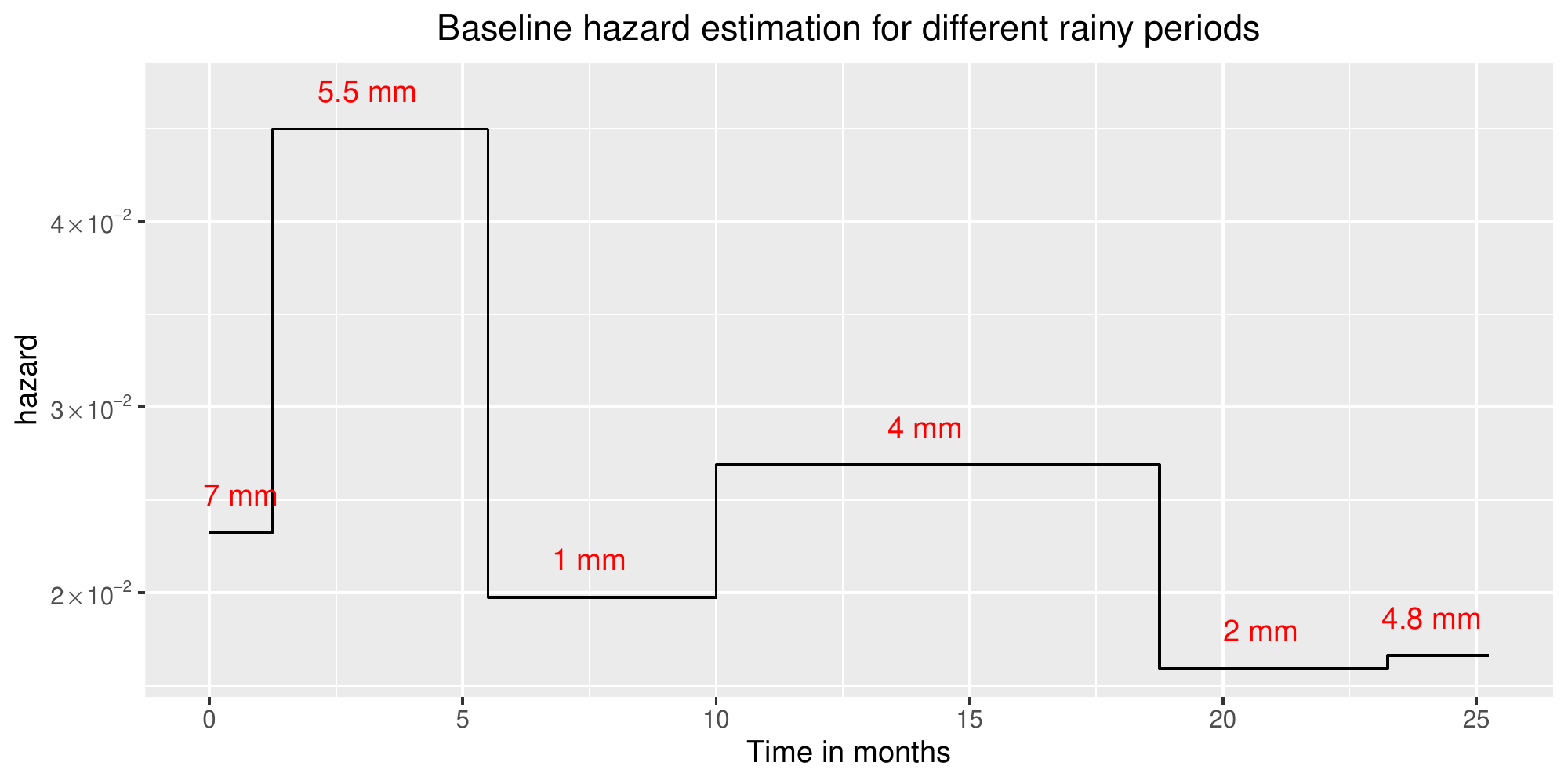}
    \caption[Hazard rate estimates for different rain patterns]{Hazard rates estimates based on univariate spatially correlated frailty model with correlation structure $\Sigma_{\text{pol}}(\rho)$ for different rain patterns. Average daily rainfall within different time periods annotated in red.}
    \label{fig:piecewise_hazard_rain}
\end{figure}

\begin{figure}[H]
    \centering
        \includegraphics[scale=0.5]{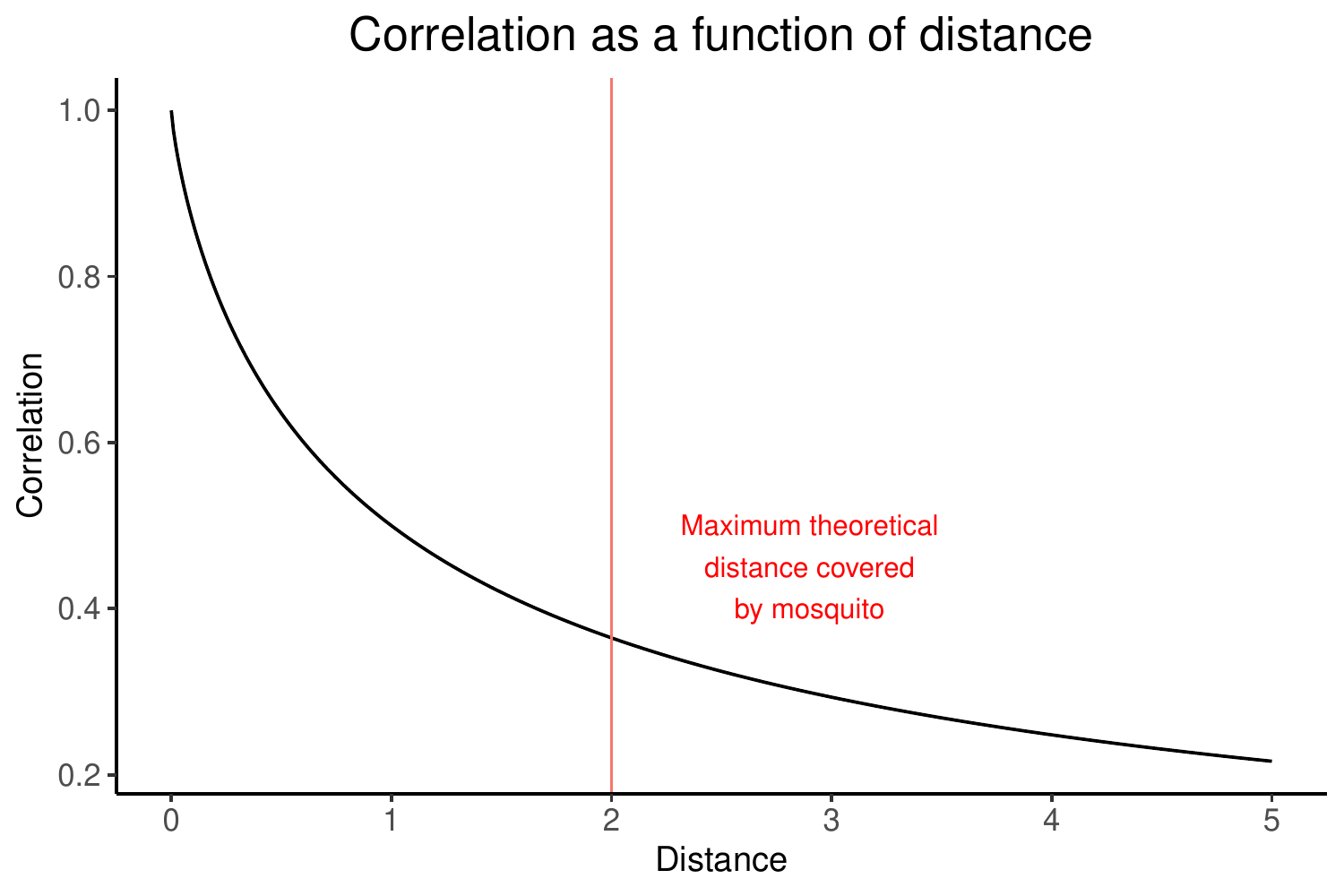}
    \caption{Correlation values $\Sigma_{\text{pol}}(\hat\rho)$ as a function of distance for $\hat\rho=0.81$.}
    \label{fig:graph_malaria_correlation_S4}
\end{figure}

The estimation results are presented in the fourth column of Table \ref{tab:marginal_shared_correlated}. We compute the model standard errors as the square root of the inverse of the Fisher Information Matrix. We refer to Appendix \ref{appendix:algorithm_calculations} for more details on the computation of the Fisher Information Matrix. The hazard ratio for the distance effect equals 1.10 (95$\%$ CI: [0.85;1.41]) and does not differ significantly from 1. Two of the other covariates had no significant effect on time to malaria, with hazard ratios equal to 0.99 (95$\%$ CI: [0.87;1.11]) and 1.03 (95$\%$ CI: [0.97;1.09]) for gender and roof type respectively. The hazard ratio for children aged between 3-7 does not differ significantly from 1 with a hazard ratio equal to 0.95 (95$\%$ CI: [0.81;1.09]). However, the hazard ratio for children older than 7 is equal to 1.42 (95$\%$ CI: [1.12;1.80]) which implies a higher malaria risk of about 42 $\%$. The piecewise baseline hazard with specific parameter estimates is depicted in Figure \ref{fig:piecewise_hazard_rain}. The malaria risk seems to be highest during or just after periods of heavy rainfall. The limited number of observations in the last interval makes the estimate of the last baseline component untrustworthy. We expect the estimate $\hat{h}_6$ to be higher had data been available for a few more weeks. With respect to the spatial correlation structure, the variance estimate $\hat{\sigma}^2$  was equal to 0.43 and $\hat{\rho}$  was equal to 0.81 with respective standard errors 0.09 and 0.14. In Figure \ref{fig:graph_malaria_correlation_S4}, we give a graphical representation of how correlation between the children evolves with respect to the distance between them imposed by these estimates.

\subsection{Comparison with results obtained with alternative models}

We analyze the malaria data using a marginal model and a shared frailty model with the frailty at the village level. Both models are defined such that the baseline hazard function is piecewise constant with cut-points determined following rainfall data as in the previous section. The results are presented in Table \ref{tab:marginal_shared_correlated}. The marginal model estimates are obtained by fitting the data to a piecewise constant proportional hazards model and then correcting for the variances of the parameters using the grouped jackknife method (\cite{wu1986jackknife}, \cite{lipsitz1996jackknife}). The hazard ratio for the distance to dam effect does not differ significantly from 1 in both models. As for children older than 7, the marginal model suggests about 46 $\%$ higher malaria risk while the shared frailty model suggests about 38 $\%$ higher malaria risk. Both models found no significant effect for gender or roof. 

\begin{table}
\begin{center}
\caption{Estimates and 95 $\%$ confidence intervals in square brackets of hazard ratios and parameters estimated in the marginal model, shared frailty model with village clustering and univariate spatially correlated frailty model. HR: Hazard ratio. MM: Marginal model, SFM: Shared frailty model, USCFM: Univariate spatially correlated frailty model.\label{tab:marginal_shared_correlated}}
\begin{tabular}{|l|l|l|l|}
\hline
 Parameters & MM & SFM & USCFM \\
\hline
$HR_{\text{sex}}$ & 0.98 [0.90;1.08] & 0.97 [0.89;1.05] & 0.98 [0.87;1.11] \\
$HR_{\text{age}}$ 3-7 & 0.91 [0.76;1.09] & 1.03 [0.95;1.11] & 0.94 [0.81;1.09] \\
 \hspace{11mm} $>$ 7 & $1.46^{*}$ [1.12;1.90] & $1.38^{*}$ [1.13;1.68] & $1.42^{*}$ [1.12;1.80] \\
$HR_{\text{d}}$ & 1.05 [0.95;1.17] & 0.94 [0.81;1.10] & 1.10 [0.85;1.41] \\
$HR_{\text{roof}}$ & 0.93 [0.82;1.06] & 1.00 [0.89;1.12] & 1.03 [0.97;1.09] \\
\hline
$h_1 $ & 0.0156 [0.0116;0.0196] & 0.0167 [0.0102;0.0233] & 0.0233 [0.0169;0.0297] \\
$h_2 $ & 0.0405 [0.0272;0.0534] & 0.0441 [0.0275;0.0609] & 0.0450 [0.0283;0.0617] \\
$h_3 $ & 0.0110 [0.0068; 0.0151] & 0.0123 [0.0074;0.0173] & 0.0198 [0.0138;0.0256] \\
$h_4 $ & 0.0185 [0.0127;0.0243] & 0.0214 [0.0136;0.0292] & 0.0269 [0.0209;0.0327] \\
$h_5 $ & 0.0063 [0.0038;0.0089] & 0.0076 [0.0045;0.0107] & 0.0159 [0.0131;0.0189] \\
$h_6 $ & 0.0070 [0.0044;0.0097] & 0.0085 [0.0049;0.0122] & 0.0166 [0.0126;0.0206] \\
\hline
$\sigma^2$ & $\times$ & 0.43 [0.35;0.52] & 0.43 [0.25;0.61] \\
\hline
$\rho$ & $\times$ & $\times$ & 0.81 [0.54;1.08] \\
\hline
\end{tabular}
\vspace{1ex}
\end{center}
{\hspace{8mm} * $\text{p-value} < 0.05$ \par}
\end{table}

\section{Discussion}

When analyzing time to event data of several subjects, it is of paramount importance to accommodate for spatial correlation at the subject level. This is even more crucial for infectious diseases where transmission occurs at the subject level. The univariate spatially correlated frailty model introduced in this paper is, to the best of our knowledge, the first such model to incorporate spatial correlation at the subject level for time to event data. Regarding model parameters estimation, it has been demonstrated through  the theoretical proof of convergence of the adapted SAEM-MCMC algorithm and the simulation results for reasonably small datasets as encountered in practice that proper estimates can be obtained for all model parameters, including those modelling the spatial correlation.

Previous approaches rather introduce spatial correlation between clusters of subjects. In many circumstances, however, and certainly in the case study on malaria analysed in this paper, the delineations of the clusters are often based on administrative rather than physical boundaries. Infectious agents are not halted, however, by such administrative
boundaries, but by distance or other physical barriers between subjects. Modelling correlation at the subject level allows to bypass such artificial cluster delimitation choices and instead include accurate physical characteristics between subjects in the model.

Moreover, it is demonstrated in the simulation studies that not taking into account spatial correlation has a serious impact on the parameter estimates, leading to serious bias, and should
thus not be neglected. Furthermore, using the villages as clusters in the marginal and shared frailty models to analyse the malaria data set has serious impact on some of the parameter estimates. The main practical objective of the malaria study was to establish the association between the distance from the dam and malaria incidence. Although no significant effect of distance was found, regardless what model was used, the direction of the effect changes. In the shared frailty model, the incidence decreases with increasing distance, whereas in the marginal and univariate spatially correlated frailty models the incidence increases with increasing distance from the dam. Although the result from the shared frailty model is the more expected one, it has serious problems as described in \cite{getachew2013} due to confounding between the distance covariate and the cluster frailty. We emphasize that the proposed univariate spatially correlated frailty model does not have this confounding problem.

Besides, the SAEM-MCMC algorithm implemented here can easily deal with other correlation structures or baseline hazard functions allowing for a wide class of  applications. Indeed, other more complex correlation structures could offer further information on the spatial correlation. Moreover one could do without any parametric modeling of the baseline hazard function and estimate the regression and frailty parameters based on the integrated partial likelihood as well, based on \cite{ajmal2019convergent}.

Spatial survival analysis is a relatively new research topic that will become more relevant with the increasing availability of geographical data. The recent COVID-19 outbreak and various tools based on GPS tracking come to mind, where contact tracing is obviously done at a subject level. The univariate spatially correlated frailty model, having shown its benefits for modelling spatially correlated malaria data, is an appropriate model to take into account the data structure of many other infectious diseases.

\appendix

\section{Appendix A}
\label{appendix:proof_theorem}

\textbf{Proof of Theorem \ref{thm:adaptative_SAEM_MCMC}}

First, we state the classical assumptions \textbf{(F3)-\textbf{(F5)}} corresponding to assumptions \textbf{(M3)}-\textbf{(M5)} of \cite{dely99} required to prove the almost sure convergence of EM like algorithms:

\textbf{(F3)} The function $\Bar{s}: \Theta \rightarrow \Sset$ defined as:

\begin{equation*}
    \Bar{s} (\theta) = \int_{\mathbb{R}^{N}} \Sborel (\frailtyv) \pi_{\theta} (\frailtyv | \bX, \bdelta) d \frailtyv
\end{equation*}

 is continuously differentiable on $\Theta$.

\textbf{(F4)} The function $l: \Theta \rightarrow \mathbb{R}$ defined as the marginal extended log-likelihood

\begin{align*}
    l(\theta) = \text{log} \int_{\mathbb{R}^{N}} L_{\text{comp}}(\theta;\bX,\bdelta,\frailtyv) d \frailtyv
\end{align*}
is continuously differentiable on $\Theta$ and 
\begin{equation*}
    \partial_{\theta} \int_{\mathbb{R}^{N}} L_{\text{comp}}(\theta;\bX,\bdelta,\frailtyv) d \frailtyv = \int_{\mathbb{R}^{N}} \partial_{\theta} \  L_{\text{comp}}(\theta;\bX,\bdelta,\frailtyv) d \frailtyv
\end{equation*}

\textbf{(F5)} There exists a function $\hat{\theta}: \Sset \rightarrow \Theta$ such that:
$$\forall \ s \in \Sset, \forall \ \theta \in \Theta, \ L(\hat{\theta}(s),s) \geq L(\theta,s)$$

where $L: \Sset \times \Theta \rightarrow \mathbb{R}$ is defined as

\begin{equation}
    L(\theta,s) = \text{exp} (- \Psi (\theta) + \langle s, \Phi (\theta) \rangle )
\end{equation}

Moreover, the function $\hat{\theta}$ is continuously differentiable on $\Sset$.

We first apply Theorem $5.5$ of \cite{moulines05} to prove the convergence of the sequence $(s_k)_k$ and check therefore the required assumptions. As detailed in \cite{moulines05}, drift assumptions \textbf{(DRI)} imply assumptions \textbf{(A2-A3)}  by Proposition 6.1. Thus we can apply Theorem $5.5$ of \cite{moulines05}. This results in the sequence $(s_k)_k$ generated by Algorithm $\mathcal{A}$ satisfying $\lim_k d(s_k, \Sset_0)=0$ where
$\Sset_0= \lbrace s \in \Sset, \langle \nabla w(s),h(s)\rangle = 0 \rbrace $. We recall that we choose a piecewise baseline function $h_0$ and we assume that $\frailtyv$ follows a multivariate normal distribution. In addition to those model hypotheses and assumptions \textbf{(F1)}, \textbf{(F2)}, we have that assumptions \textbf{(M1)} and \textbf{(M2)} of \cite{dely99} hold in our case. It suffices to show that $\Psi$ and $\Phi$ are twice continuously differentiable to satisfy assumption \textbf{(M2)}. This is guaranteed by assumptions \textbf{(F1)}, \textbf{(F2)}, the choices of the frailty distribution and of the baseline hazard function $h_0$. As shown in equation (\ref{eq:Lcomp_spatial_exponential}), we can write the complete likelihood in exponential form which fulfils assumption \textbf{(M1)}. Following the lines of the proof of Lemma 2 of \cite{dely99}, we get that $\lim_k d(\theta_k, \mathcal{L})=0$. The proof of Theorem \ref{thm:adaptative_SAEM_MCMC} is therefore complete.

\section{Appendix B}
\label{appendix:algorithm_calculations}
\begin{enumerate}

\item[1.] \textbf{Equations for the Maximization step of Algorithm $\mathcal{A}$}

Let us first recall the complete log-likelihood of the data~:
\begin{eqnarray*}
\log L_{\text{comp}} (\theta;\bX,\bdelta,\frailtyv) &=& \sum_{i=1}^{N} \Bigg( \dobs \Big( \log \big(\sum_{m=1}^{M} h_m \mathbbm{1}_{ [ \tau_{m-1}, \tau_m [ } (\yobs) \big) + \\
&& \desfixe^t \beta + \frailty \Big) - H_0(\yobs) \exp(\desfixe^t \beta + \frailty) \Bigg)  + \log f(\frailtyv) \nonumber
\end{eqnarray*}
where the cumulative hazard function $H_0(\yobs) = \sum_{m=1}^M h_m (\tau_m - \tau_{m-1}) \mathbbm{1}_{[\tau_m, \infty[} (\yobs) + \sum_{m=1}^M (\yobs - \tau_{m-1}) h_m \mathbbm{1}_{[\tau_{m-1},\tau_m[} (\yobs)$. 

We differentiate the complete log-likelihood with respect to each parameter to obtain the necessary equations to update the parameter estimates in Algorithm $\mathcal{A}$. 
\medskip 
\justify
\textbf{Updating parameters $(h_m)_{1 \leq m \leq M}$}
\justify
Differentiating the complete log-likelihood with respect to $h_m$, we obtain: 

\begin{align}
    \frac{\partial \log L_{\text{comp}} (\theta;\bX,\bdelta,\frailtyv)}{\partial h_m} &= \frac{\sum_{i=1}^{N} \Delta_{i} \mathbbm{1}_{ [ \tau_{m-1}, \tau_m [ } (\yobs)}{h_m} \nonumber \\  
    &- \sum_{i=1}^{N} \exp (\desfixe^t \beta + b_{i})  ( (\tau_m - \tau_{m-1}) \mathbbm{1}_{[\tau_m, \infty[} (\yobs) \nonumber \\
    &+  (\yobs - \tau_{m-1}) \mathbbm{1}_{[\tau_{m-1},\tau_m[} (\yobs) )
\end{align}

We obtain an analytic expression for the update of $h_m$:

\begin{equation*}
    \hat{h}_m = \frac{\sum_{i=1}^{N} \Delta_{i} \mathbbm{1}_{ [ \tau_{m-1}, \tau_m [ } (\yobs)}
      {\sum_{i=1}^{N} \exp (\desfixe^t \beta + b_{i})  \left( (\tau_m - \tau_{m-1}) \mathbbm{1}_{[\tau_m, \infty[} (\yobs) + (\yobs - \tau_{m-1}) \mathbbm{1}_{[\tau_{m-1},\tau_m[} (\yobs)  \right)}
\end{equation*}

\justify
\textbf{Updating parameters $\beta$}
\justify
Differentiating the complete log-likelihood with respect to $\beta$, we obtain:

\begin{align}
\begin{split}
    \frac{\partial \log L_{\text{comp}} (\theta;\bX,\bdelta,\frailtyv)}{\partial \beta} = \sum_{i=1}^{N} \Big( \dobs \desfixe^t - \desfixe^t H_0(\yobs) \text{ exp } (\desfixe^t \beta + \frailty)  \Big)
\end{split}
\end{align}

\begin{align}
\begin{split}
    \frac{\partial^2 \log L_{\text{comp}} (\theta;\bX,\bdelta,\frailtyv)}{\partial \beta^2} = - \sum_{i=1}^{N} \desfixe \desfixe^t H_0(\yobs) \text{ exp } (\desfixe^t \beta + \frailty)
\end{split}
\end{align}

The Newton Raphson method is used to update the values of parameter $\beta$. 
\justify
\textbf{Updating parameter $\sigma^2$}
\justify
Only the last term of the log-likelihood depends on the parameter $\sigma^2$. Differentiating the log-likelihood with respect to $\sigma^2$ we get:

\begin{align}
\begin{split}
    \frac{\partial \log L_{\text{comp}} (\theta;\bX,\bdelta,\frailtyv)}{\partial \sigma^2} = - \frac{N}{2 \sigma^2} + \frac{1}{2 (\sigma^2)^{2}} \frailtyv^t \Sigma (\rho)^{-1} \frailtyv
\end{split}
\end{align}

We obtain an analytic expression to update the parameter $\sigma^2$:

\begin{equation}
    \hat{\sigma}^2 = \frac{1}{N} \frailtyv^t \Sigma (\hat{\rho})^{-1} \frailtyv
\end{equation}

\justify
\textbf{Updating parameter $\rho$}
\justify
We now differentiate the log-likelihood with respect to $\rho$ leading to: 

\begin{align}
    \frac{\partial \log L_{\text{comp}} (\theta;\bX,\bdelta,\frailtyv)}{\partial \rho} &= - \frac{1}{2} \text{tr} \Big( \Sigma (\rho)^{-1} \frac{\partial}{\partial \rho} \Sigma (\rho) \Big) + \frac{1}{2} (\sigma^2)^{-1} \frailtyv^t \Sigma (\rho)^{-1} \frac{\partial}{\partial \rho} \Sigma (\rho) \Sigma (\rho)^{-1} \frailtyv
\end{align}

The values of the parameter $\rho$ are updated using  a gradient descent method.

\item[2.] \textbf{Computation of the marginal likelihood}

We recall that the frailty $\frailtyv$ is assumed to follow a multivariate normal distribution parameterized by $\Gamma$. A numerical approximation of the marginal likelihood is computed based on the parameter estimates $\hat{\theta} = ((\hat{h}_m)_{1 \leq 1 \leq M},\hat{\beta},\hat{\sigma}^2,\hat{\rho})$. We simulate $C$ independent realizations of $(\frailtyv_c)_{1 \leq c \leq C}$ following a multivariate normal distribution with mean zero and covariance matrix $\hat{\sigma}^2 \Sigma(\hat{\rho})$. The marginal likelihood is then computed based on a Monte Carlo sum as follows:
$$\hat{L}_{\text{marg}} (\theta;\bX,\bdelta) = \frac{1}{C} \sum_{c=1}^{C} L_{\text{cond}} (\hat{\theta};\bX,\bdelta|\frailtyv_c)$$
where 
\begin{align*}
\label{eq:complete_likelihood_spatial}
L_{\text{cond}} (\theta;\bX,\bdelta|\frailtyv) = \prod_{i=1}^{N} \Bigg( \frac{\Big( \sum_{m=1}^{M} h_m \mathbbm{1}_{ [ \tau_{m-1}, \tau_m [ } (\yobs) \exp(\desfixe^t \beta + \frailty) \Big)^{\dobs}}{\exp(H_0(\yobs) \exp(\desfixe^t \beta + \frailty))} \Bigg)
\end{align*}

The law of large numbers (\cite{van2000asymptotic}) ensures that the Monte Carlo sum converges to the marginal likelihood as $C \rightarrow \infty$. The bigger the number of realizations $C$, the better the quality of the approximation. We compute $\hat{L}_{\text{marg}}$ for increasing values of $C$. The values obtained are compared and once they are of the same order, the quality of the approximation is determined to be sufficient.

\item[3.] \textbf{Computation of the Fisher information matrix}

We obtain an estimate of the Fisher information matrix through the observed Fisher information matrix $I_{obs}(\theta) = - \partial_{\theta}^2 \text{log} L_{\text{marg}} (\theta;\bX,\bdelta)$ (\cite{andersen97}). Using Louis's missing information principle (\cite{louis}), we express the matrix $I_{obs}(\theta)$ as:

\begin{align*}
I_{obs}(\theta) &= - \mathbb{E}_{\theta} \big( \partial_{\theta}^2 \log L_{\text{comp}} (\theta;\bX,\bdelta, \frailtyv) | \bX,\bdelta \big) - \\
&\text{Cov}_{\theta} \big( \partial_{\theta} \log L_{\text{comp}} (\theta;\bX,\bdelta, \frailtyv) | \bX,\bdelta \big)
\end{align*}
where  $\mathbb{E}_{\theta}$ and $\text{Cov}_{\theta}$ denote respectively the expectation and the covariance under the posterior distribution $\pi_{\theta}$ of the frailty.

We approximate the quantity $I_{obs}(\theta)$ by a Monte Carlo sum based on the realizations of the Markov chain generated in the algorithm having as stationary distribution the posterior distribution $\pi_{\theta}$. After a  burn-in period, we use the remaining $L$ realizations $(\frailtyv_L)_{1 \leq l \leq L}$ of the Markov chain to compute the following quantity:

\begin{align*}
\hat{I}_L(\hat{\theta}) &= - \frac{1}{L} \sum_{l=1}^L  \partial_{\theta}^2 \log L_{\text{comp}} (\hat{\theta};\bX,\bdelta, \frailtyv_l) \\
& - \frac{1}{L} \sum_{l=1}^L \big( \partial_{\theta} \log L_{\text{comp}} (\hat{\theta};\bX,\bdelta,\frailtyv_l)  \partial_{\theta} \log L_{\text{comp}} (\hat{\theta};\bX,\bdelta,\frailtyv_l)^t \big) \\
& + \frac{1}{L^2} \bigg( \sum_{l=1}^L \partial_{\theta} \log L_{\text{comp}} (\hat{\theta};\bX,\bdelta,\frailtyv_l) \bigg)  \bigg( \sum_{l=1}^L \partial_{\theta} \log L_{\text{comp}} (\hat{\theta};\bX,\bdelta,\frailtyv_l) \bigg)^t
\end{align*}

The ergodic theorem  guarantees the convergence of the quantity $\hat{I}_L(\hat{\theta})$ to the observed Fisher information matrix $I_{obs}(\hat{\theta})$ as $L$ goes to infinity (\cite{meyn}). In addition to the derivatives calculated to compute the M-step of Algorithm $\mathcal{A}$, we also compute the following second and cross derivatives:  

\begin{align*}
    \frac{\partial^2 \log L_{\text{comp}} (\theta;\bX,\bdelta,\frailtyv)}{\partial h_m^2} &= - \frac{\sum_{i=1}^{N} \Delta_{i} \mathbbm{1}_{ [ \tau_{m-1}, \tau_m [ } (\yobs)}{h_m^2} 
\end{align*}

\begin{align*}
\begin{split}
    \frac{\partial^2 \log L_{\text{comp}} (\theta;\bX,\bdelta,\frailtyv)}{\partial (\sigma^2)^2} = \frac{N}{2 (\sigma^2)^2} - \frac{1}{(\sigma^2)^{3}} \frailtyv^t \Sigma (\rho)^{-1} \frailtyv
\end{split}
\end{align*}

\begin{align*}
    \frac{\partial^2 \log L_{\text{comp}} (\theta;\bX,\bdelta,\frailtyv)}{\partial \rho^2} &= - \frac{1}{2} \text{tr} \Big( - A(\rho) + \Sigma (\rho)^{-1} \frac{\partial^2}{\partial \rho^2} \Sigma (\rho) \Big) \nonumber \\ 
    &+ \frac{1}{2} (\sigma^2)^{-1} \frailtyv^t \Big( \lbrace - A(\rho) \nonumber \\
    &+ \Sigma (\rho)^{-1} \frac{\partial^2}{\partial \rho^2} \Sigma (\rho) \rbrace \Sigma (\rho)^{-1} - A(\rho) \  \Sigma (\rho)^{-1} \Big) \frailtyv
\end{align*}

\begin{align*}
    \frac{\partial^2 \log L_{\text{comp}} (\theta;\bX,\bdelta,\frailtyv)}{\partial \beta \partial h_m} &= - \sum_{i=1}^{N}  \desfixe^t \text{ exp } (\desfixe^t \beta + \frailty) \\
    & \times  \left( (\tau_m - \tau_{m-1}) \mathbbm{1}_{[\tau_m, \infty[} (\yobs) + (\yobs - \tau_{m-1}) \mathbbm{1}_{[\tau_{m-1},\tau_m[} (\yobs)  \right) 
\end{align*}

\begin{align*}
    \frac{\partial^2 \log L_{\text{comp}} (\theta;\bX,\bdelta,\frailtyv)}{\partial \rho \partial \sigma^2} &= -
     \frac{1}{2(\sigma^2)^2 } \frailtyv^t \Sigma (\rho)^{-1} \frac{\partial}{\partial \rho} \Sigma (\rho) \Sigma (\rho)^{-1} \frailtyv
\end{align*}

\end{enumerate}

\bibliographystyle{Chicago}
\bibliography{biblio_frailty}

\begin{thebibliography}{}

\bibitem[\protect\citeauthoryear{{Allassonni\`ere}, {Kuhn}, and
  {Trouv\'e}}{{Allassonni\`ere} et~al.}{2010}]{stephkuhn}
{Allassonni\`ere}, S., E.~{Kuhn}, and A.~{Trouv\'e} (2010).
\newblock Construction of bayesian deformable models via a stochastic
  approximation algorithm: A convergence study.
\newblock {\em Bernoulli\/}~{\em 16}, 641--678.

\bibitem[\protect\citeauthoryear{{Andersen}, {Klein}, {Knudsen}, and {Tabanera
  y Palacios}}{{Andersen} et~al.}{1997}]{andersen97}
{Andersen}, P., J.~{Klein}, K.~{Knudsen}, and R.~{Tabanera y Palacios} (1997).
\newblock Estimation of variance in cox's regression model with shared gamma
  frailties.
\newblock {\em Biometrics\/}~{\em 53}, 1475--1484.

\bibitem[\protect\citeauthoryear{{Andrieu}, {Moulines}, and
  {Priouret}}{{Andrieu} et~al.}{2005}]{moulines05}
{Andrieu}, C., E.~{Moulines}, and P.~{Priouret} (2005).
\newblock Stability of stochastic approximation under verifiable conditions.
\newblock {\em SIAM J. Control Optim\/}~{\em 44}, 283--312.

\bibitem[\protect\citeauthoryear{Banerjee}{Banerjee}{2016}]{banerjee2016spatial}
Banerjee, S. (2016).
\newblock Spatial data analysis.
\newblock {\em Annual review of public health\/}~{\em 37}, 47--60.

\bibitem[\protect\citeauthoryear{Banerjee, Wall, and Carlin}{Banerjee
  et~al.}{2003}]{banerjee2003frailty}
Banerjee, S., M.~M. Wall, and B.~P. Carlin (2003).
\newblock Frailty modeling for spatially correlated survival data, with
  application to infant mortality in minnesota.
\newblock {\em Biostatistics\/}~{\em 4\/}(1), 123--142.

\bibitem[\protect\citeauthoryear{Belay, Kifle, Goshu, Gran, Yewhalaw,
  Duchateau, and Frigessi}{Belay et~al.}{2017}]{belay2017}
Belay, D.~B., Y.~G. Kifle, A.~T. Goshu, J.~M. Gran, D.~Yewhalaw, L.~Duchateau,
  and A.~Frigessi (2017).
\newblock Joint bayesian modeling of time to malaria and mosquito abundance in
  ethiopia.
\newblock {\em BMC infectious diseases\/}~{\em 17\/}(1), 415.

\bibitem[\protect\citeauthoryear{Brostr{\"o}m}{Brostr{\"o}m}{2012}]{brostrom2012eha}
Brostr{\"o}m, G. (2012).
\newblock eha: Event history analysis. r package version 2.0-7.

\bibitem[\protect\citeauthoryear{Delyon, Lavielle, and Moulines}{Delyon
  et~al.}{1999}]{dely99}
Delyon, B., M.~Lavielle, and E.~Moulines (1999).
\newblock Convergence of a stochastic approximation version of the {E}{M}
  algorithm.
\newblock {\em Ann. Statist.\/}~{\em 27\/}(1), 94--128.

\bibitem[\protect\citeauthoryear{Getachew, Janssen, Yewhalaw, Speybroeck, and
  Duchateau}{Getachew et~al.}{2013}]{getachew2013}
Getachew, Y., P.~Janssen, D.~Yewhalaw, N.~Speybroeck, and L.~Duchateau (2013).
\newblock Coping with time and space in modelling malaria incidence: a
  comparison of survival and count regression models.
\newblock {\em Statistics in medicine\/}~{\em 32\/}(18), 3224--3233.

\bibitem[\protect\citeauthoryear{Haario, Saksman, Tamminen, et~al.}{Haario
  et~al.}{2001}]{haario2001adaptive}
Haario, H., E.~Saksman, J.~Tamminen, et~al. (2001).
\newblock An adaptive metropolis algorithm.
\newblock {\em Bernoulli\/}~{\em 7\/}(2), 223--242.

\bibitem[\protect\citeauthoryear{{Kuhn} and {Lavielle}}{{Kuhn} and
  {Lavielle}}{2004}]{KuhnLavielle04}
{Kuhn}, E. and M.~{Lavielle} (2004).
\newblock Coupling a stochastic approximation version of em with an mcmc
  procedure.
\newblock {\em ESAIM: Probability and Statistics\/}~{\em 8}, 115--131.

\bibitem[\protect\citeauthoryear{Li and Ryan}{Li and Ryan}{2002}]{louise02}
Li, Y. and L.~Ryan (2002).
\newblock Modeling spatial survival data using semiparametric frailty models.
\newblock {\em Biometrics\/}~{\em 58\/}(2), 287--297.

\bibitem[\protect\citeauthoryear{Lin}{Lin}{2012}]{lin2012}
Lin, P.-S. (2012).
\newblock Analysis of spatial frailty models by a weighted estimating equation.
\newblock {\em Journal of Statistical Planning and Inference\/}~{\em 142\/}(6),
  1436--1444.

\bibitem[\protect\citeauthoryear{Lipsitz and Parzen}{Lipsitz and
  Parzen}{1996}]{lipsitz1996jackknife}
Lipsitz, S.~R. and M.~Parzen (1996).
\newblock A jackknife estimator of variance for cox regression for correlated
  survival data.
\newblock {\em Biometrics\/}~{\em 52\/}(1), 291--298.

\bibitem[\protect\citeauthoryear{{Louis}}{{Louis}}{1982}]{louis}
{Louis}, T. (1982).
\newblock Finding the observed information matrix when using the em algorithm.
\newblock {\em J. Roy. Statist. Soc. Ser. B\/}~{\em 44}, 226--233.

\bibitem[\protect\citeauthoryear{Meyn and Tweedie}{Meyn and
  Tweedie}{2012}]{meyn}
Meyn, S.~P. and R.~L. Tweedie (2012).
\newblock {\em Markov chains and stochastic stability}.
\newblock Springer Science \& Business Media.

\bibitem[\protect\citeauthoryear{Oodally, Duchateau, and Kuhn}{Oodally
  et~al.}{2019}]{ajmal2019convergent}
Oodally, A., L.~Duchateau, and E.~Kuhn (2019).
\newblock Convergent stochastic algorithm for parameter estimation in frailty
  models using integrated partial likelihood.
\newblock {\em arXiv preprint arXiv:1909.07056\/}.

\bibitem[\protect\citeauthoryear{Ripatti, Larsen, and Palmgren}{Ripatti
  et~al.}{2002}]{ripatti2002maximum}
Ripatti, S., K.~Larsen, and J.~Palmgren (2002).
\newblock Maximum likelihood inference for multivariate frailty models using an
  automated monte carlo em algorithm.
\newblock {\em Lifetime Data Analysis\/}~{\em 8\/}(4), 349--360.

\bibitem[\protect\citeauthoryear{Self and Liang}{Self and
  Liang}{1987}]{self1987asymptotic}
Self, S.~G. and K.-Y. Liang (1987).
\newblock Asymptotic properties of maximum likelihood estimators and likelihood
  ratio tests under nonstandard conditions.
\newblock {\em Journal of the American Statistical Association\/}~{\em
  82\/}(398), 605--610.

\bibitem[\protect\citeauthoryear{Van~der Vaart}{Van~der
  Vaart}{2000}]{van2000asymptotic}
Van~der Vaart, A.~W. (2000).
\newblock {\em Asymptotic statistics}, Volume~3.
\newblock Cambridge university press.

\bibitem[\protect\citeauthoryear{{World Health Organization}}{{World Health
  Organization}}{2018}]{worldmalaria2018}
{World Health Organization} (2018).
\newblock World malaria report 2018.

\bibitem[\protect\citeauthoryear{Wu et~al.}{Wu et~al.}{1986}]{wu1986jackknife}
Wu, C.-F.~J. et~al. (1986).
\newblock Jackknife, bootstrap and other resampling methods in regression
  analysis.
\newblock {\em the Annals of Statistics\/}~{\em 14\/}(4), 1261--1295.

\bibitem[\protect\citeauthoryear{Yewhalaw, Getachew, Tushune, Kassahun,
  Duchateau, Speybroeck, et~al.}{Yewhalaw et~al.}{2013}]{yewhalaw2013}
Yewhalaw, D., Y.~Getachew, K.~Tushune, W.~Kassahun, L.~Duchateau,
  N.~Speybroeck, et~al. (2013).
\newblock The effect of dams and seasons on malaria incidence and anopheles
  abundance in ethiopia.
\newblock {\em BMC infectious diseases\/}~{\em 13\/}(1), 161.

\end{thebibliography}
\end{document}